\documentclass[twocolumn,useAMS,usenatbib]{mnras}
\usepackage{natbib}
\usepackage{fmtcount}
\usepackage{threeparttable}
\usepackage{graphicx,subfig}
\usepackage{multirow}
\usepackage{url}
\usepackage{amssymb}

\title[1987A-like SN 2009mw]{Optical photometry and spectroscopy of the 1987A-like supernova 2009mw}
\author[K. Tak\' ats et al.]{K. Tak\' ats$^{1,2}$\thanks{E-mail: ktakats@gmail.com}, G. Pignata$^{2,1}$, M. Bersten$^{3,4,5}$, M.~L. Rojas Kaufmann$^3$,  \newauthor J.~P. Anderson$^{6}$, G. Folatelli$^{3,4,5}$, M. Hamuy$^{7,1}$, M. Stritzinger$^{8}$, J.~B.~Haislip$^{9}$, \newauthor A.~P.~LaCluyze$^{9}$, J.~P.~Moore$^{9}$ and D.~Reichart$^{9}$ \\
$^1$Millennium Institute of Astrophysics,Vicu\~na Mackenna 4860, 7820436 Macul, Santiago, Chile \\
$^2$Departamento de Ciencias Fisicas, Universidad Andres Bello, Avda. Republica 252, Santiago, Chile\\
$^3$Facultad de Ciencias Astron\'omicas y Geof\'isicas, Universidad Nacional de La Plata (UNLP), Paseo del Bosque S/N, B1900FWA,\\ La Plata, Argentina\\
$^4$Instituto de Astrof\'isica La Plata, IALP, CCT-CONICET-UNLP, Paseo del Bosque s/n, 1900 La Plata, Argentina\\
$^5$Kavli Institute for the Physics and Mathematics of the Universe (WPI), The University of Tokyo, Kashiwa, Chiba 277-8583, Japan\\
$^6$European Southern Observatory, Alonso de C\'ordova 3107, Casilla 19001, Santiago, Chile\\
$^7$Departamento de Astronom\'ia, Universidad de Chile, Casilla 36-D, Santiago, Chile\\
$^8$Department of Physics and Astronomy, Aarhus University, Ny Munkegade, DK-8000 Aarhus C, Denmark\\
$^{9}$University of North Carolina at Chapel Hill, Campus Box 3255, Chapel Hill, NC 27599-3255, USA\\
}

\begin{document}
\date{Accepted  Received ; in original form }

\pagerange{\pageref{firstpage}--\pageref{lastpage}} \pubyear{}

\maketitle

\label{firstpage}

\begin{abstract}

We present optical photometric and spectroscopic observations of the 1987A-like supernova (SN) 2009mw. Our $BVRI$ and $g'r'i'z'$ photometry covers 167 days of evolution, including the rise to the light curve maximum, and ends just after the beginning of the linear tail phase. We compare the observational properties of SN~2009mw with those of other SNe belonging to the same subgroup, and find that it shows similarities to several objects. The physical parameters of the progenitor and the SN are estimated via hydrodynamical modelling, yielding an explosion energy of $1$~foe, a pre-SN mass of $19\,{\rm M_{\odot}}$, a progenitor radius as $30\,{\rm R_{\odot}}$ and a $^{56}$Ni mass as $0.062\,{\rm M_{\odot}}$. These values indicate that the progenitor of SN~2009mw was a blue supergiant star, similar to the progenitor of SN~1987A. We examine the host environment of SN~2009mw and find that it emerged from a population with slightly sub-solar metallicty.

\end{abstract}


%################################### Intro
\section{Introduction}

The explosion of supernova (SN) 1987A in the Large Magellanic Cloud was a unique event that has immeasurably improved our understanding of the evolution of massive stars. It is the most studied SN, with a vast amount of literature published. It was also a high-impact event in the sense that very few similar SNe have been observed so far. These objects are core-collapse events with spectra akin to those of the very common SNe~II-P. However, while both models and observations found that SNe II-P emerge from red supergiant (RSG) stars, the progenitor of SN~1987A was a blue supergiant \citep[BSG, see][ and references therein]{arnett_87A}.

1987A-like SNe are intrinsically rare, \citet{pastorello_09E} estimated that they represent about $1-3$ per cent of all core-collapse SNe in a volume-limited sample. In fact, after SN 1987A we had to wait 11 years for the discovery of the next event, SN~1998A \citep{pastorello_98a}. Since then there have been a few more objects studied. \citet{kleiser_2000cb} published  observations of SNe~2000cb and 2005ci, though they note that SN~2000cb does not match the characteristics of either SNe~II-P or SN~1987A, its $B$ band light curve is more similar to that of SNe~II-P, while in the redder bands it shows the slow rising light curve of 1987A-like SNe, therefore its classification as 1987A-like object is questionable. \citet{taddia_87alike} presented data of SNe~2006V and 2006au, while  \citet{pastorello_09E} published SN~2009E, all of which are well-observed examples of this subgroup. \citet{pastorello_09E} compiled the known sample of 1987A-like SNe to date, 11 objects in total, including SN~1987A, but also SN~2000cb. \citet{arcavi_iau} showed the light curves of SNe~2004ek, 2004em and 2005dp that might also belong to this subgroup. \citet{taddia_87Alike_2016} analysed the data of SNe~2004ek, 2004em, 2005ci and 2005dp, and added to the sample PTF09gpn, and PTF12kso. The light curve of SN~2004ek resembles to that of SN~2000cb. \citet{kelly_snrefsdal_87A} classified SN Refsdal as a luminous 1987A-like SN. In summary, the entire available sample of 1987A-like SNe consists of maximum 17 objects, only a handful of them having well-sampled, multiband light curves and spectroscopic observations.

In this paper we report on the observations of SN~2009mw, a member of the group of 1987A-like objects.
SN~2009mw was discovered on December 23.30~UT 2009 in ESO 499-G005 by the Chilean Supernova Search \citep[CHASE, ][]{chase} project, and was classified as a Type II-P SN on December 25.2~UT \citep{09mw_felfed}. The classification spectrum suggested a normal SN II-P about a month after the explosion, however, the subsequent photometric follow-up revealed a light curve similar to that of SN~1987A.

This paper is organized as follows. In Sect.~\ref{sec_photometry} we present the photometric data and light curve analysis. In Sect.~\ref{sec_spectroscopy} we show the spectroscopic follow-up data and compare it to those of other 1987A-like SNe. We estimate the physical parameters of the progenitor and the explosion of SN~2009mw via hydrodynamical modelling in Sect.~\ref{sec_physicalparameters}, and analyse the host environment in Sect.~\ref{sec_host}. Finally, the results are summarized and discussed in Sect.~\ref{sec_discussion}.

%################################### Photometry
\section{Photometry}\label{sec_photometry}

\begin{figure}
\includegraphics[width=84mm]{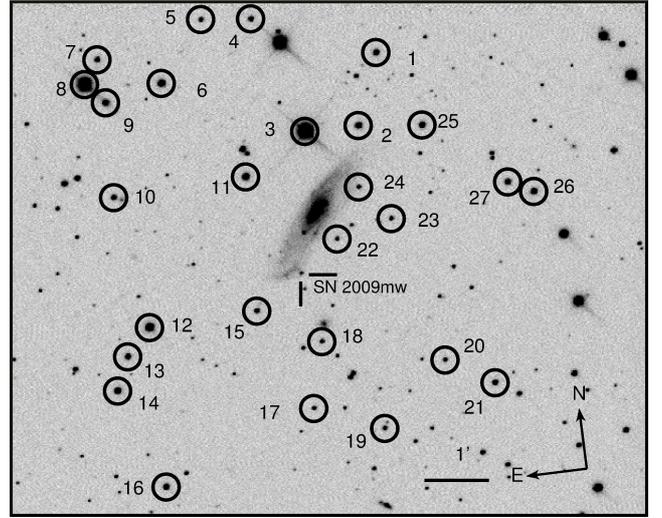}
\caption{The field of SN 2009mw taken with the PROMPT5 telescope in the $R$ band. The local sequence of stars used for the photometric calibration are marked with numbers. Their magnitudes can be found in Table \ref{localseq}.}
\label{picture}
\end{figure}

\begin{figure}
\includegraphics[width=84mm]{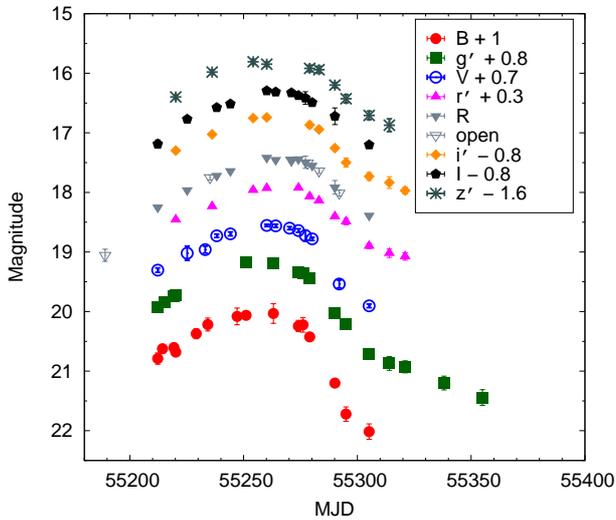}
\caption{The $BVRI$ and $g'r'i'z$ light curves of SN~2009mw}
\label{lc}
\end{figure}

Optical photometric observations of SN 2009mw were obtained with the PROMPT Telescopes in $BVRI$, $g'r'i'z'$, and open filters. The data reduction was carried out using standard {\sc iraf}\footnote{{\sc iraf} is distributed by the National Optical Astronomy Observatories, which are operated by the Association of Universities for Research in Astronomy, Inc., under the cooperative agreement with the National Science Foundation.} tasks. After the basic reduction steps (bias-substraction, overscan-correction, flat-fielding), the photometric measurments of the SN were performed with the point-spread function (PSF) fitting technique. We calibrated the photometry by observing standard fields \citep{landolt,landolt2007,sloan_std_fields} on photometric nights. With the help of these images, magnitudes for a local sequence of stars (Fig.~\ref{picture}, Table~\ref{localseq}) in the SN field were determined and used to calibrate the SN measurements (Fig.~\ref{lc}, Table \ref{lc_table}). Some of our observations were taken without filter. Since the effective wavelength without filter shaped by the CCD QE is close to the effective wavelength of the $R$ band, we decided to calibrate the unfitered instrumental magnitudes to the latter band. To compute the transformation we used a colour term estimated by observing Landolt standard stars without filter i.e. with the same setup used for the SN observations.

Our observations lasted for about 167~d after discovery, covering the evolution around the maximum  and ending right after the tail phase started. Fitting a low order polynomial around the maximum, we determined that the $B$-band maximum occured $71$~d after the discovery epoch, i.e. on $MJD=55258.8 \pm 0.4$, with an apparent magnitude of $m_B = 18.99 \pm 0.02$~mag.

\begin{figure}
\includegraphics[width=84mm]{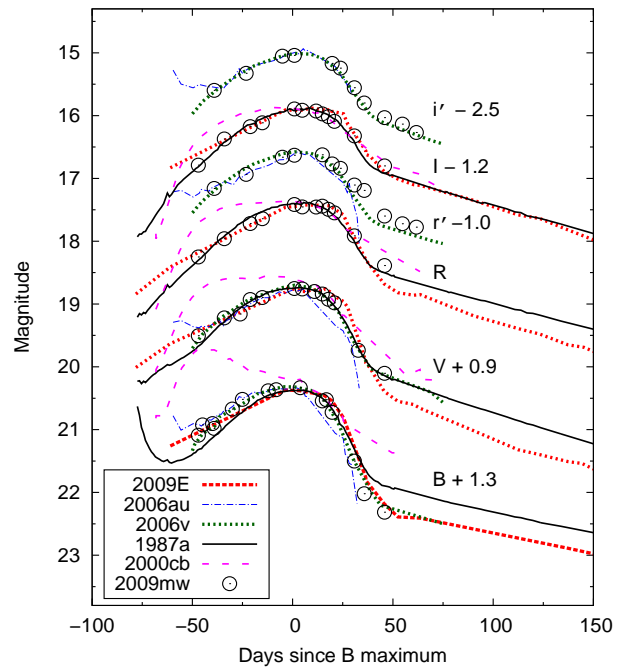}
\caption{Comparison of the shapes of the $BVRr'Ii'$ light curves of a selected sample of 1987A-like SNe to those of SN 2009mw. The light curves of the comparison SNe were shifted vertically to match the magnitudes of SN 2009mw at maximum and horizontally to match the epoch of the maximum. Since the $B$ band light curve of SN~2000cb does not show the broad delayed maximum like the others, it was shifted (somewhat arbitrarily) by $76$~days relative to the the explosion date for the sake of the comparison.}
\label{lc_comp}
\end{figure}

In Fig.~\ref{lc_comp} we compare the $BVRr'Ii'$ light curve shapes of several 1987A-like SNe with those of SN~2009mw. Some relevant parameters of the SNe that we use for comparison throughout the paper can be found in Table~\ref{otherSN}. In Fig.~\ref{lc_comp} the light curves of the comparison SNe are shifted vertically to match the peak magnitudes of SN 2009mw. This figure shows that the light curve shapes of the different objects are quite similar, only those of SN 2000cb differ significantly, showing flatter curves in $VRI$ bands and missing the broad, late maximum of the $B$ band curve entirely.

Since we do not have non-detections to constrain the explosion epoch of SN 2009mw, we have to rely on comparisons with other similar objects. In the small sample of this subgroup, the only SN with well-known explosion epoch is SN 1987A. The $B$ maximum of this object occurred $82.2 \pm 1.1$ days after the explosion. Another SN with some constrains on the explosion epoch is SN 2009E \citep{pastorello_09E}, that exploded $86.5 \pm 6.8$ days before the $B$ maximum. This value is quite similar to that of SN 1987A, but with higher uncertainties. We determined the explosion epoch of SN 2009mw by adopting the weighted mean of the two values, $84.3 \pm 3.5$~d, leading to the explosion epoch $MJD=55174.5$ (2009 Dec 9.5 UT), $14$~d before discovery.

\begin{table*}
\tiny
 \centering
 \begin{minipage}{\textwidth}
  \caption{$BVRI$ and $g'r'i'z'$ magnitudes of the local sequence of stars, used for photometric calibration.}
  \label{localseq}
  \begin{tabular}{@{}lcccccccccc@{}}
  \hline
ID & Ra & Dec & $B$ & $V$ & $R$ & $I$ & $g'$ & $r'$ & $i'$ & $z'$ \\
\hline
 1 &  09:47:07.942 &  -24:48:00.74  & 16.511 (0.025)  & 15.546 (0.035)  & 15.044 (0.043)  & 14.508 (0.047)  &  16.003 (0.013)  & 15.273 (0.029) &  14.979 (0.017)  &  14.815 (0.004) \\
 2 &  09:47:09.775 &  -24:49:06.95  & 15.480 (0.019)  & 15.025 (0.026)  & 14.745 (0.043)  & 14.456 (0.043)  &  15.246 (0.029)  & 14.936 (0.035) &  14.847 (0.014)  &  14.819 (0.009)  \\
 3 &  09:47:13.471 &  -24:49:06.13  & 13.053 (0.032)  & 11.575 (0.021)  &                 &                 &  12.256 (0.020)  & 11.081 (0.032) &  10.571 (0.089)  &  10.177 (0.012)  \\
 4 &  09:47:16.186 &  -24:47:14.58  & 16.741 (0.035)  & 16.176 (0.027)  & 15.805 (0.053)  & 15.438 (0.026)  &  16.418 (0.024)  & 16.029 (0.030) &  15.863 (0.026)  &  15.778 (0.017)  \\
 5 &  09:47:19.600 &  -24:47:09.37  & 17.502 (0.050)  & 16.793 (0.016)  & 16.391 (0.046)  & 15.978 (0.039)  &  17.122 (0.033)  & 16.601 (0.043) &  16.373 (0.049)  &  16.327 (0.024)  \\
 6 &  09:47:22.843 &  -24:48:04.46  & 14.843 (0.018)  & 14.402 (0.030)  & 14.129 (0.050)  & 13.831 (0.031)  &  14.610 (0.022)  & 14.316 (0.042) &  14.217 (0.018)  &  14.196 (0.013)  \\
 7 &  09:47:27.022 &  -24:47:34.86  & 16.733 (0.034)  & 15.989 (0.022)  & 15.585 (0.068)  & 15.190 (0.032)  &  16.321 (0.026)  & 15.802 (0.036) &  15.606 (0.018)  &  15.518 (0.011)  \\
 8 &  09:47:28.092 &  -24:47:56.56  & 12.471 (0.018)  & 11.714 (0.032)  & 11.282 (0.046)  & 10.836 (0.028)  &  12.056 (0.024)  & 11.503 (0.034) &  11.268 (0.020)  &  11.157 (0.011)  \\
 9 &  09:47:26.815 &  -24:48:16.03  & 15.809 (0.023)  & 15.203 (0.031)  & 14.872 (0.046)  & 14.490 (0.047)  &  15.486 (0.021)  & 15.066 (0.034) &  14.905 (0.022)  &  14.832 (0.007)  \\
10 &  09:47:27.083 &  -24:49:45.66  & 17.515 (0.033)  & 16.317 (0.051)  & 15.613 (0.063)  & 14.972 (0.041)  &  16.906 (0.015)  & 15.883 (0.038) &  15.462 (0.028)  &  15.223 (0.004)  \\
11 &  09:47:17.905 &  -24:49:41.74  & 15.496 (0.017)  & 14.754 (0.032)  & 14.337 (0.043)  & 13.949 (0.035)  &  15.124 (0.021)  & 14.550 (0.034) &  14.375 (0.013)  &  14.289 (0.008)  \\
12 &  09:47:25.767 &  -24:51:51.35  & 15.251 (0.014)  & 14.143 (0.034)  & 13.483 (0.052)  & 12.909 (0.039)  &  14.695 (0.013)  & 13.746 (0.031) &  13.382 (0.019)  &  13.194 (0.019)  \\
13 &  09:47:27.512 &  -24:52:16.16  & 16.123 (0.029)  & 15.521 (0.025)  & 15.208 (0.056)  & 14.843 (0.047)  &  15.802 (0.012)  & 15.400 (0.027) &  15.260 (0.021)  &  15.210 (0.020)  \\
14 &  09:47:28.495 &  -24:52:46.94  & 15.623 (0.020)  & 14.947 (0.030)  & 14.603 (0.045)  & 14.229 (0.043)  &  15.260 (0.011)  & 14.804 (0.031) &  14.645 (0.019)  &  14.578 (0.006)  \\
15 &  09:47:18.305 &  -24:51:48.53  & 17.077 (0.039)  & 16.337 (0.023)  & 15.949 (0.048)  & 15.536 (0.051)  &  16.672 (0.015)  & 16.176 (0.043) &  15.987 (0.022)  &  15.871 (0.041)  \\
16 &  09:47:26.020 &  -24:54:22.44  & 16.790 (0.034)  & 15.719 (0.027)  & 15.127 (0.067)  & 14.541 (0.038)  &  16.248 (0.006)  & 15.383 (0.046) &  15.024 (0.042)  &  14.809 (0.010)  \\
17 &  09:47:15.264 &  -24:53:26.23  & 17.343 (0.036)  & 16.978 (0.061)  & 16.798 (0.046)  & 16.507 (0.039)  &  17.123 (0.012)  & 16.957 (0.037) &  16.895 (0.021)  &  16.922 (0.058)  \\
18 &  09:47:14.135 &  -24:52:24.69  & 17.767 (0.050)  & 17.128 (0.038)  & 16.767 (0.045)  & 16.411 (0.059)  &  17.453 (0.045)  & 16.991 (0.029) &  16.834 (0.018)  &  16.736 (0.014)  \\
19 &  09:47:10.597 &  -24:53:53.18  & 17.627 (0.059)  & 16.872 (0.028)  & 16.473 (0.059)  & 16.056 (0.064)  &  17.217 (0.025)  & 16.688 (0.029) &  16.478 (0.011)  &  16.399 (0.021)  \\
20 &  09:47:05.893 &  -24:52:56.25  & 18.063 (0.129)  & 17.122 (0.039)  & 16.651 (0.039)  & 16.167 (0.050)  &  17.563 (0.039)  & 16.873 (0.012) &  16.632 (0.026)  &  16.476 (0.042)  \\
21 &  09:47:02.678 &  -24:53:23.33  & 16.929 (0.009)  & 16.035 (0.021)  & 15.495 (0.063)  & 14.967 (0.033)  &  16.461 (0.011)  & 15.737 (0.034) &  15.424 (0.020)  &  15.251 (0.021)  \\
22 &  09:47:12.206 &  -24:50:50.50  & 17.757 (0.037)  & 17.016 (0.076)  & 16.647 (0.020)  & 16.213 (0.033)  &  17.382 (0.025)  & 16.851 (0.020) &  16.682 (0.024)  &  16.553 (0.051)  \\
23 &  09:47:08.336 &  -24:50:37.78  & 17.774 (0.058)  & 16.992 (0.062)  & 16.580 (0.082)  & 16.149 (0.035)  &  17.356 (0.014)  & 16.804 (0.024) &  16.592 (0.035)  &  16.472 (0.033)  \\
24 &  09:47:10.292 &  -24:50:04.39  & 17.922 (0.084)  & 17.261 (0.035)  & 16.893 (0.046)  & 16.489 (0.043)  &  17.596 (0.055)  & 17.099 (0.041) &  16.896 (0.025)  &  16.794 (0.056)  \\
25 &  09:47:05.428 &  -24:49:13.87  & 15.829 (0.020)  & 15.253 (0.032)  & 14.927 (0.046)  & 14.588 (0.046)  &  15.525 (0.028)  & 15.119 (0.033) &  15.003 (0.016)  &  14.954 (0.020)  \\
26 &  09:46:58.392 &  -24:50:29.08  & 15.694 (0.026)  & 15.088 (0.029)  & 14.751 (0.048)  & 14.394 (0.040)  &  15.370 (0.037)  & 14.944 (0.035) &  14.800 (0.019)  &  14.741 (0.009)  \\
27 &  09:47:00.077 &  -24:50:17.13  & 16.764 (0.028)  & 15.718 (0.030)  & 15.072 (0.046)  & 14.507 (0.039)  &  16.228 (0.031)  & 15.318 (0.038) &  14.979 (0.017)  &  14.797 (0.023)  \\
\hline
\end{tabular}
\end{minipage}
\end{table*}

\begin{table*}
\tiny
 \centering
 \begin{minipage}{\textwidth}
  \caption{$BVRI$ and $g'r'i'z'$ magnitudes of SN 2009mw, obtained with the PROMPT telescopes.}
  \label{lc_table}
  \begin{tabular}{@{}lccccccccc@{}}
  \hline
Date & MJD & $B$ & $V$ & $R$ & $I$ & $g'$ & $r'$ & $i'$ & $z'$ \\
\hline
2009/12/23$^a$ &  55188.3 & & & 19.124 (0.079) & & & & &  \\
2009/12/24$^a$ &  55189.1 & & & 19.055 (0.102) & & & & &   \\
2009/12/27$^a$ &  55192.2 & & & 18.893 (0.065) & & & & &   \\
2010/01/10$^a$ &  55206.3 & & & 18.457 (0.057) & & & & &   \\
2010/01/11$^a$ &  55207.8 & & & 18.380 (0.052) & & & & &   \\
2010/01/15$^a$ &  55211.2 & & & 18.226 (0.044) & & & & &   \\
2010/01/16     &  55212.2 & 19.789 (0.095) & 18.605 (0.037) & 18.250 (0.030) & 17.988 (0.043) & 19.121 (0.070) & & &  \\
2010/01/18     &  55214.4 & 19.625 (0.072) & & & & & & &   \\
2010/01/19     &  55215.2 & & & & & 19.042 (0.058) &  &  &  \\
2010/01/23     &  55219.3 & 19.604 (0.061) & & & & 18.950 (0.059) & & & \\
2010/01/24     &  55220.2 & 19.681 (0.062) & & & & 18.921 (0.050) & 18.159 (0.026) & 18.097 (0.026) & 17.999 (0.0369  \\
2010/01/29     &  55225.2 & & 18.321 (0.034) & 17.960 (0.028) & 17.571 (0.027) & & & & \\
2010/02/02     &  55229.2 & 19.372 (0.084) & & & & & & &  \\
2010/02/06     &  55233.1 & & 18.260 (0.058) & & & & & &  \\
2010/02/07     &  55234.2 & 19.218 (0.048) & & & & & & &  \\
2010/02/08$^a$ &  55235.2 & & & 17.750 (0.042) & & & & &  \\
2010/02/09     &  55236.1 & & & & & & 17.935 (0.023) & 17.826 (0.027) & 17.581 (0.038) \\
2010/02/11     &  55238.1 & & 18.028 (0.029) & 17.718 (0.024) & & & & & \\
2010/02/17     &  55244.1 & & 17.996 (0.032) & 17.641 (0.026) & 17.315 (0.021) & & & & \\
2010/02/20     &  55247.1 & 19.081 (0.056) & & & & & & & \\
2010/02/24     &  55251.1 & 19.064 (0.030) & & & & 18.373 (0.021) & & & \\
2010/02/27     &  55254.2 & & & & & & 17.662 (0.027) & 17.553 (0.030) & 17.407 (0.042) \\
2010/03/05     &  55260.2 & & 17.854 (0.019) & 17.416 (0.019) & 17.093 (0.022) & & 17.630 (0.026) & 17.541 (0.029) & 17.447 (0.037) \\
2010/03/08     &  55263.1 & 19.033 (0.044) & & & & 18.382 (0.033) & & &  \\
2010/03/09     &  55264.1 & & 17.861 (0.028) & 17.453 (0.024) & 17.114 (0.020) & & & & \\
2010/03/15     &  55270.2 & & 17.900 (0.025) & & & & & & \\
2010/03/16$^a$ &  55271.0 & & & 17.455 (0.039) & & & & & \\
2010/03/16     &  55271.1 & & & 17.456 (0.021) & 17.127 (0.023) & & & & \\
2010/03/19     &  55274.1 & 19.243 (0.046) & 17.94. (0.029) & 17.441 (0.027) & 17.173 (0.022) & 18.537 (0.039) & 17.624 (0.026) & & \\
2010/03/21     &  55276.1 & 19.223 (0.065) & & & & 18.555 (0.038) & & & \\
2010/03/22     &  55277.1 & & 18.025 (0.040) & 17.493 (0.023) & 17.216 (0.039) & & & & \\
2010/03/23$^a$ &  55278.1 & & & 17.506 (0.051) & & & & & \\
2010/03/24     &  55279.0 & 19.425 (0.049) & & & & 18.645 (0.036) & 17.770 (0.039) & 17.668 (0.051) & 17.519 (0.055) \\
2010/03/25     &  55280.1 & & 18.081 (0.027) & 17.548 (0.022) & 17.291 (0.020) & & & & \\
2010/03/28$^a$ &  55283.1 & & & 17.638 (0.044) & & & 17.839 (0.025) & 17.743 (0.045) & 17.541 (0.059) \\
2010/04/04     &  55290.1 & 20.201 (0.035) & & 17.915 (0.115) & 17.522 (0.140) & 19.234 (0.033) & 18.105 (0.032) & 18.056 (0.037) & 17.799 (0.044) \\
2010/04/06     &  55292.0 & & 18.838 (0.066) & & & & & & \\
2010/04/06$^a$ &  55292.0 & & & 18.011 (0.052) & & & & & \\
2010/04/09     &  55295.0 & 20.721 (0.129) & & & & 19.410 (0.068) & 18.191 (0.053) & 18.299 (0.076) & 18.028 (0.069) \\
2010/04/19     &  55305.1 & 21.015 (0.129) & 19.204 (0.049) & 18.387 (0.032) & 18.001 (0.033) & 19.906 (0.078) & 18.600 (0.044) & 18.532 (0.066) & 18.308 (0.072) \\
2010/04/21     &  55307.0 & & 19.112 (0.086) & & 18.045 (0.052) & & & & \\
2010/04/27     &  55313.0 & & 19.330 (0.053) & 18.562 (0.035) & 18.158 (0.039) & & & & \\
2010/04/28     &  55314.0 & & & & & 20.070 (0.116) & 18.721 (0.071) & 18.635 (0.099) & 18.472 (0.108) \\
2010/05/05     &  55321.0 & & & & & 20.128 (0.101) & 18.776 (0.062) & 18.769 (0.055) & \\
2010/05/22     &  55338.0 & & & & & 20.400 (0.116) & & & \\
2010/06/04     &  55351.0 & & 19.754 (0.059) & & & & & & \\
2010/06/08     &  55355.0 & & & 18.876 (0.042) & & 20.643 (0.131) & & & \\
\hline
\end{tabular}
 \begin{tablenotes}
       \item[a]{$^a$Open filter.}

     \end{tablenotes}
\end{minipage}
\end{table*}

\begin{table*}
 \centering
 \begin{minipage}{\textwidth}
  \caption{The explosion epoch, the epoch of the $B$ maximum, the maximum absolute magnitude in $B$ band, the reddening and the distance of the SNe used for comparison throughout the paper.}
  \label{otherSN}
  \begin{tabular}{@{}lcccccc@{}}
  \hline
  \hline
SN & $t_{exp}$ & $t_{max,B}$ & M$_{max,B}$ & $E(B-V)$ & $D$ & Ref. \\
 & (MJD) & (MJD) & (mag) & (mag) & (Mpc) & \\
\hline
1987A & 46849.3 (0.0) & 46931.5 & -14.75 (0.01) & 0.19 & 0.05 (0.005) & 1,2 \\
1998A &  50801.0 (4.0) & 50890.0 & -15.22 (0.20) & 0.12  & 29.2 (2.0) & 3 \\
2000cb & 51655.5 (4.1) & & & 0.11 & 32.2 (8.0) & 4 \\
2006V & 53747.5 (4.0) & 53823.2 & -16.19 (0.15) & 0.03 & 74.5 (5.0) & 5 \\
2006au & 53793.5 (9.0) & 53865.0 & -16.02 (0.14) & 0.31 & 47.4 (3.2) & 5 \\
2009E & 54832.0 (3.0) & 54918.5 & -14.53 (0.17) & 0.04 & 29.9 (2.2) & 6 \\
\hline
\end{tabular}
 \begin{tablenotes}
       \item[a]{References: (1) \citet{menzies_87A}, (2) \citet{catchpole_87A_I}, (3)~\citet{pastorello_98a}, (4) \citet{kleiser_2000cb}, (5) \citet{taddia_87alike}, (6) \citet{pastorello_09E}.}
       \item[b]{Note, that we use the value $H_0=72\,{\rm km}\,{\rm s}^{-1}\,{\rm Mpc}^{-1}$ throughout the paper.}
     \end{tablenotes}
\end{minipage}
\end{table*}

%############################### reddening
\subsection{Reddening and colour curves}

The galactic reddening in the direction of SN~2009mw is $E(B-V)_{MW}=0.054 \pm 0.001$~mag \citep{reddening_recalib}. In the spectra (Section~\ref{sec_spectroscopy}) there are no visible Na\,{\sc i}~D absorption lines, the usual tracers of host galaxy extinction. Therefore we assume the host galaxy extinction component to be negligible, and adopt the galactic component as the total reddening towards the SN.

\begin{figure}
\includegraphics[width=84mm]{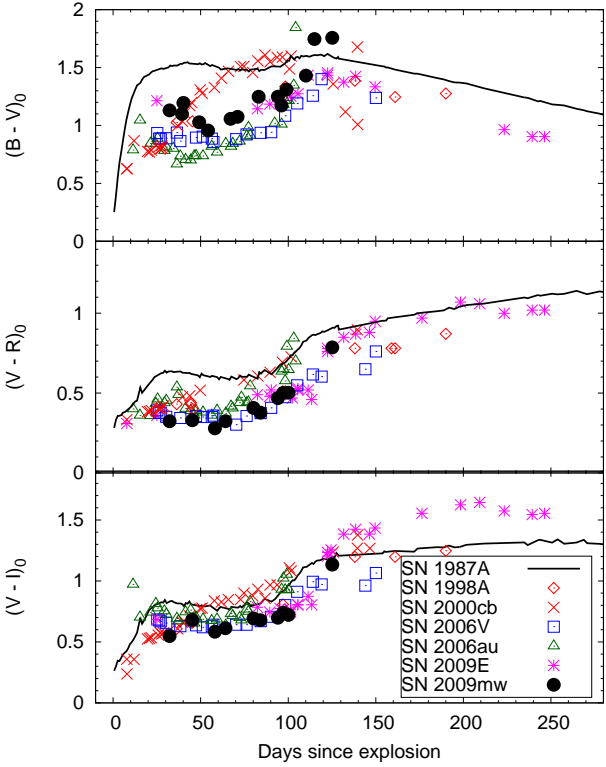}
\caption{The reddening-corrected $(B-V)_0$, $(V-R)_0$ and $(R-I)_0$ colour curves of SN~2009mw, in comparison with the colour curves of other 1987A-like SNe. Note, that the $RI$ magnitudes of SNe 2006V and 2006au were transformed from the $ri$ bands (see text).}
\label{colour}
\end{figure}

In Fig.~\ref{colour} we show the $(B-V)_0$, $(V-R)_{0}$ and $(V-I)_0$ colours corrected with $E(B-V)=0.054$~mag. Since the epochs of the $B$ band photometry differ from the others, we interpolated the $V$ band light curve to the epochs of the $B$ band observations in order to calculate the $(B-V)_0$ colour curve. In Fig.~\ref{colour} we also compare the colours to those of other 1987A-like objects. The reddening values of these SNe can be found in Table~\ref{otherSN}. In the case of SNe 2006V and 2006au $ri$ band magnitudes are available instead of $RI$ band observations \citep{taddia_87alike}. We included these objects in the colour comparison by transforming the $ri$ magnitudes to the Johnson system using the formulae of \citet{sloan_to_johnson}. Since these equations were calibrated for stars, first we tested them by converting $r'i'$
magnitudes of SN 2009mw to $RI$ magnitudes and comparing the result with measured values. We found that they agree within $1 \sigma$. Encouraged by this,  we converted the $ri$ magnitudes of SNe 2006V and 2006au to the $RI$ bands and used these values for the colour comparison.

Fig.~\ref{colour} shows the colour comparison between SN~2009mw and the other SNe in our sample. SN~1987A has redder colours during its pre-maximum evolution than most of the sample, but the difference seems to disappear at later times. The $B-V$ colour of SN~2000cb keeps reddening steadily during the first $\sim 80$ days of evolution. There is some scatter among the rest of the SNe in the $(B-V)_0$ colour, while in the other two colours these objects are similar. The $(V-R)_0$ and $(V-I)_0$ colours of all the SNe, except those of SN~2000cb, are almost constant until about $110-120$ days after explosion ($\sim 30$ days after maximum) then suddenly the SNe become redder. This change probably marks the end of the recombination (plateau) phase. At phases later than 120 d after explosion the $V-I$ colour of SN~2009E is redder than that of the other SNe, though not all have such late time observations available.

%################################distance

\subsection{Distance and bolometric light curve}\label{sec_dist_bol}

There are several distance measurments in the literature for the host galaxy of SN~2009mw, ESO~499-G005, all using the Tully-Fisher (TF) method. \citet{TF1_ngc1559} obtained the Malmquist-corrected TF distance of $44.69 \pm 8.5$ Mpc. \citet{JHK-TF} used the TF method in $JHK$ band and obtained the average distance of $57.93 \pm 10.6$~Mpc. \citet{springob_2009} found the Malquist-corrected TF distance as $44.29 \pm 8.13$~Mpc. The redshift distance is $57 \pm 4$~Mpc\footnote{NED; \url{http://ned.ipac.caltech.edu/}}. In this paper we adopt the weighted mean of these distances, $D=50.95 \pm 4.07$~Mpc ($\mu=33.54 \pm 0.17$~mag).

The $B$ band peak absolute magnitude of SN~2009mw using the obtained distance is $M_{B,max}=-14.77 \pm 0.17$~mag. This value is among the lowest in our sample of 1987A-like SNe (Table~\ref{otherSN}), it is similar to that of SN~1987A and only slightly higher than that of SN~2009E.

\begin{figure}
\includegraphics[width=84mm]{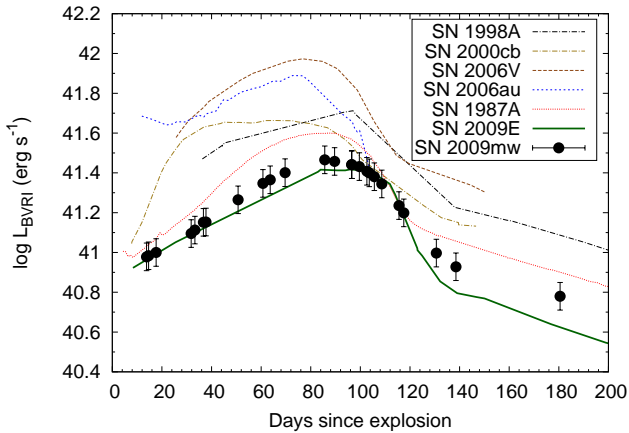}
\caption{$BVRI$ quasi-bolometric light curve of SN~2009mw and comparison to other 1987A-like SNe.}
\label{bollum}
\end{figure}

In Fig.\ref{bollum} we compare the $BVRI$ quasi-bolometric luminosities of 1987A-like SNe to that of SN 2009mw. We used the same method for all of the SNe to calculate these luminosities, by converting the $BVRI$ magnitudes to fluxes and integrating them using Simpson's rule. The distances of the SNe used for comparison can be found in Table \ref{otherSN}.
The luminosity of SN~2009mw is similar to that of SN~2009E and somewhat lower than that of SN~1987A.

%################################## Spectro
\section{Spectroscopy}\label{sec_spectroscopy}

\begin{figure}
\includegraphics[width=84mm]{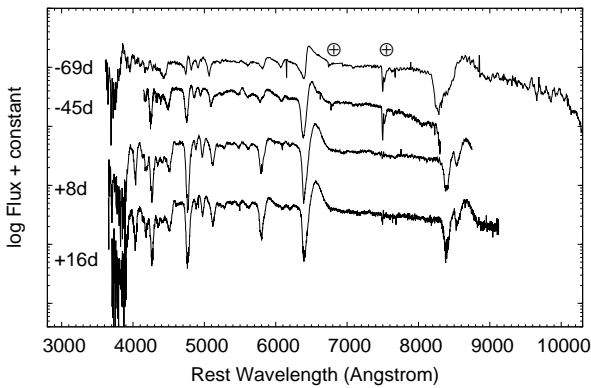}
\caption{The spectral sequence of SN 2009mw. The positions of strong telluric features are marked with the symbol $\oplus$.}
\label{sp_evol}
\end{figure}

Four optical spectra were taken of SN~2009mw with the 8.1-m Gemini South Telescope (+GMOS), the 4.1-m Southern Astrophysical Research Telescope (SOAR+Goodman)  and the 2.5-meter Ir\'en\'ee du Pont telescope (+WFCCD) (see Table \ref{logofsp}). The reduction, extraction and calibration of the spectra were carried out using standard {\sc iraf} tasks. In the case of the Gemini-S data, we used the tasks in the {\sc gemini} package in {\sc iraf}. Note, that the wavelength calibration of the first spectrum is somewhat uncertain, since we had to use the night sky lines for calibration. The reduced spectra are shown in Fig.~\ref{sp_evol}.

\begin{figure*}
\includegraphics[width=\textwidth]{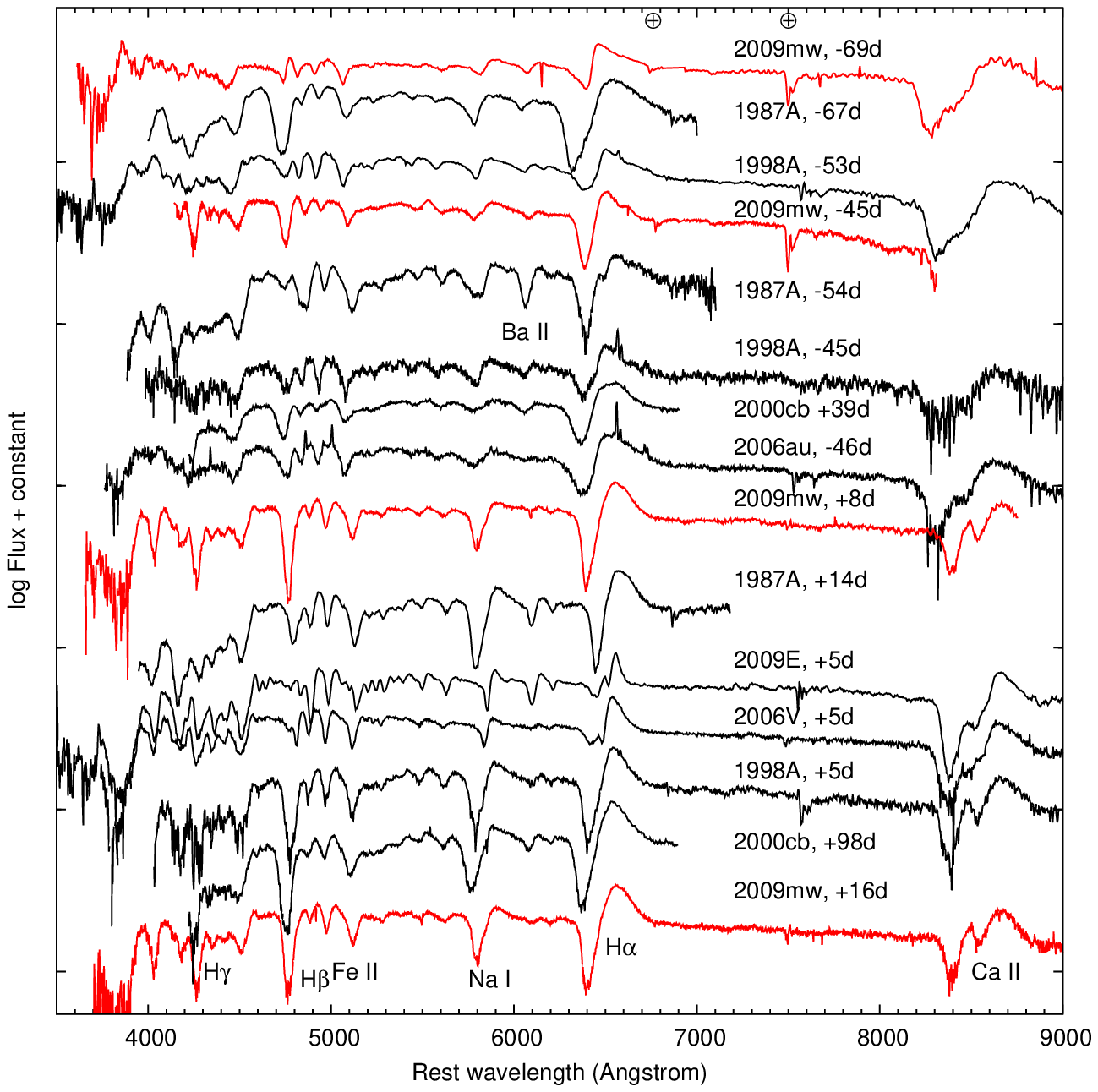}
\caption{Comparison of the spectra of SN~2009mw with those of other 1987A-like SNe, namely SNe 1987A \citep{1987A_spectra}, 1998a \citep{pastorello_98a}, 2000cb \citep{kleiser_2000cb}, 2006V, 2006au \citep{taddia_87alike} and 2009E \citep{pastorello_09E}.}
\label{sp_comp}
\end{figure*}

The spectra show features that are typically present in SNe II-P during the recombination phase, such as the H\,{\sc i} Balmer series, Na\,{\sc ii}, Ca\,{\sc ii}, Ba\,{\sc ii}, Fe\,{\sc ii}, Sc\,{\sc ii}, Ti\,{\sc ii} (Fig~\ref{sp_comp}).  During the observed period the same features stay present, while the Na\,{\sc i} and H\,{\sc i} lines strengthen significantly.

In Fig~\ref{sp_comp} we compare the spectra of SN~2009mw to those of other 1987A-like SNe at similar phases. The spectra of SN~2009mw are the most similar to those of SNe~1998A and 2006au, mainly because of their more prevalent H$\beta$ absorption features and very weak Ba\,{\sc ii} lines. The Ba\,{\sc ii} features are quite strong in the spectra of SNe 1987A and 2009E. Strong Ba\,{\sc ii} features often appear in the spectra of subluminous SNe II-P and their appearance are usually explained invoking temperature effects \citep[see e.g.][]{turatto_ba, pastorello_09E, takats2014}.

\begin{figure*}
\centering
\subfloat{
        \label{}
        \includegraphics[width=0.45\textwidth]{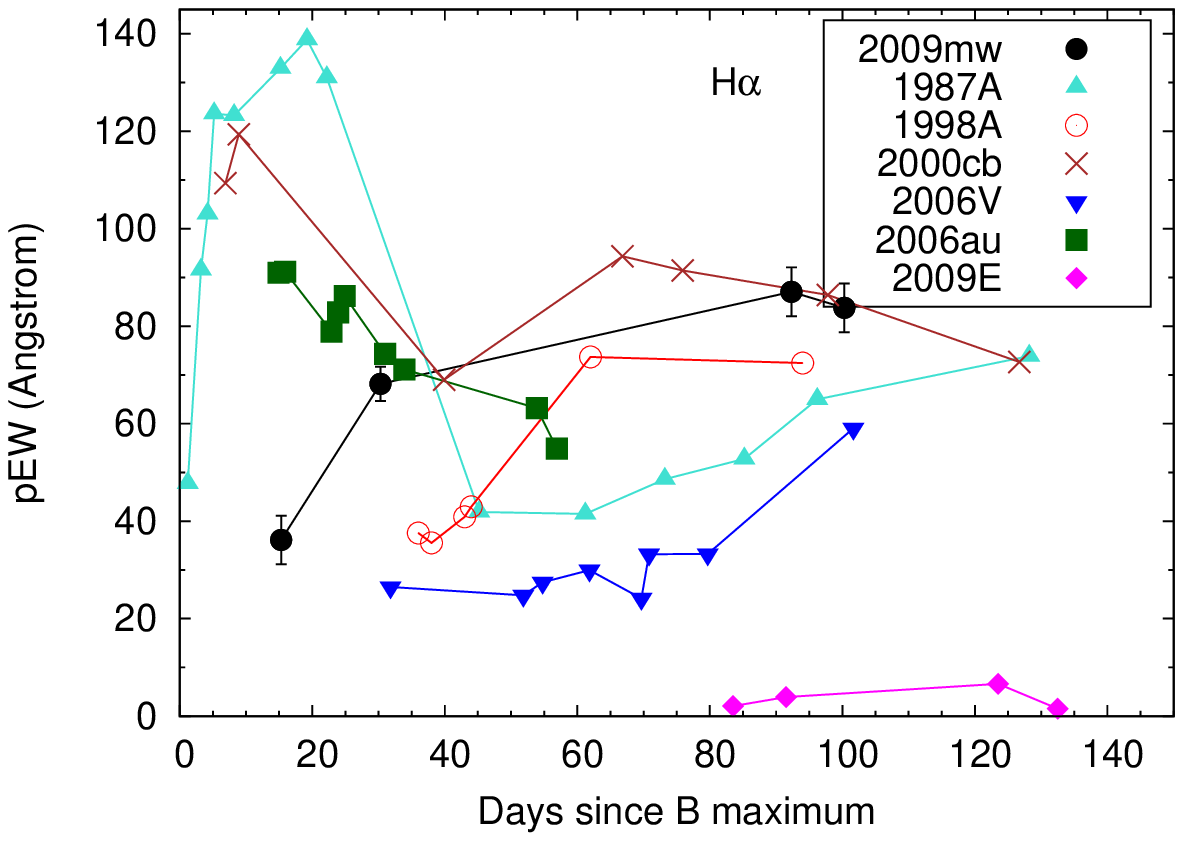} }
\subfloat{
        \label{}
        \includegraphics[width=0.45\textwidth]{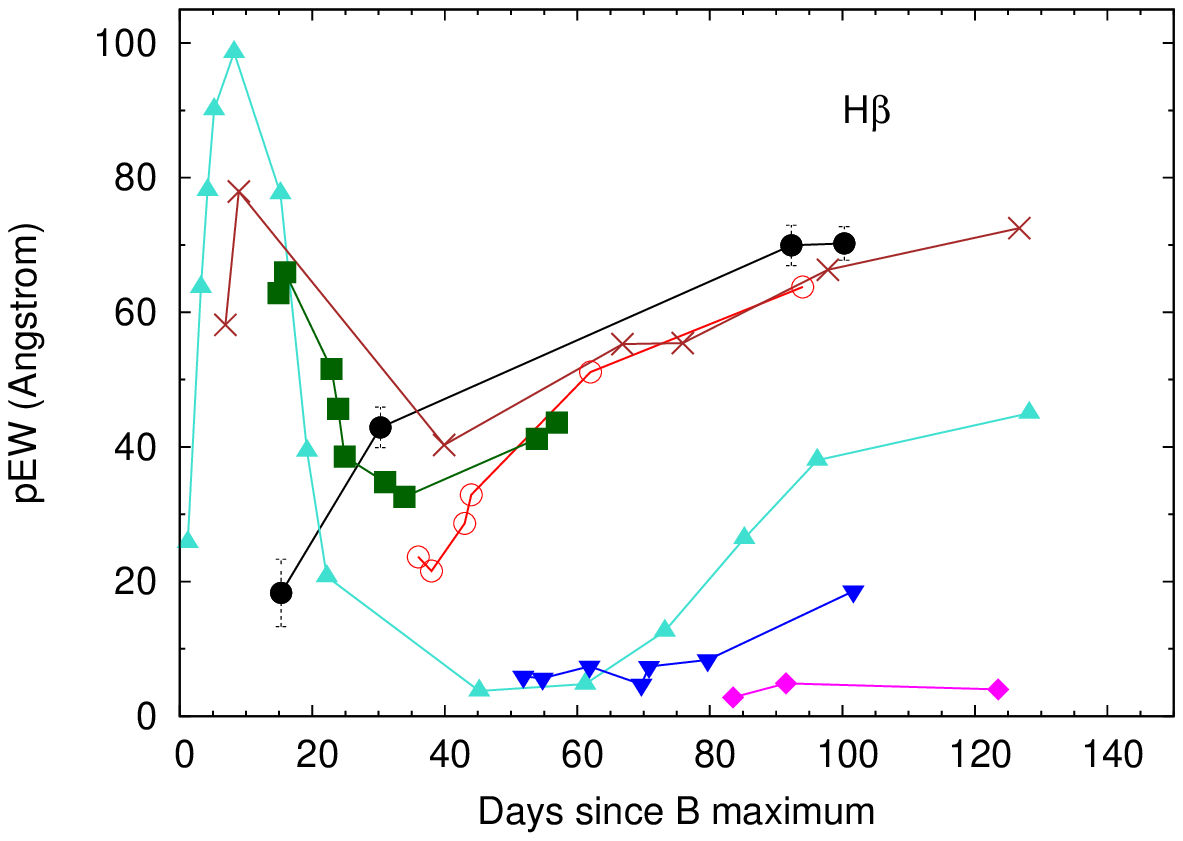} }\\
\subfloat{
        \label{}
        \includegraphics[width=0.45\textwidth]{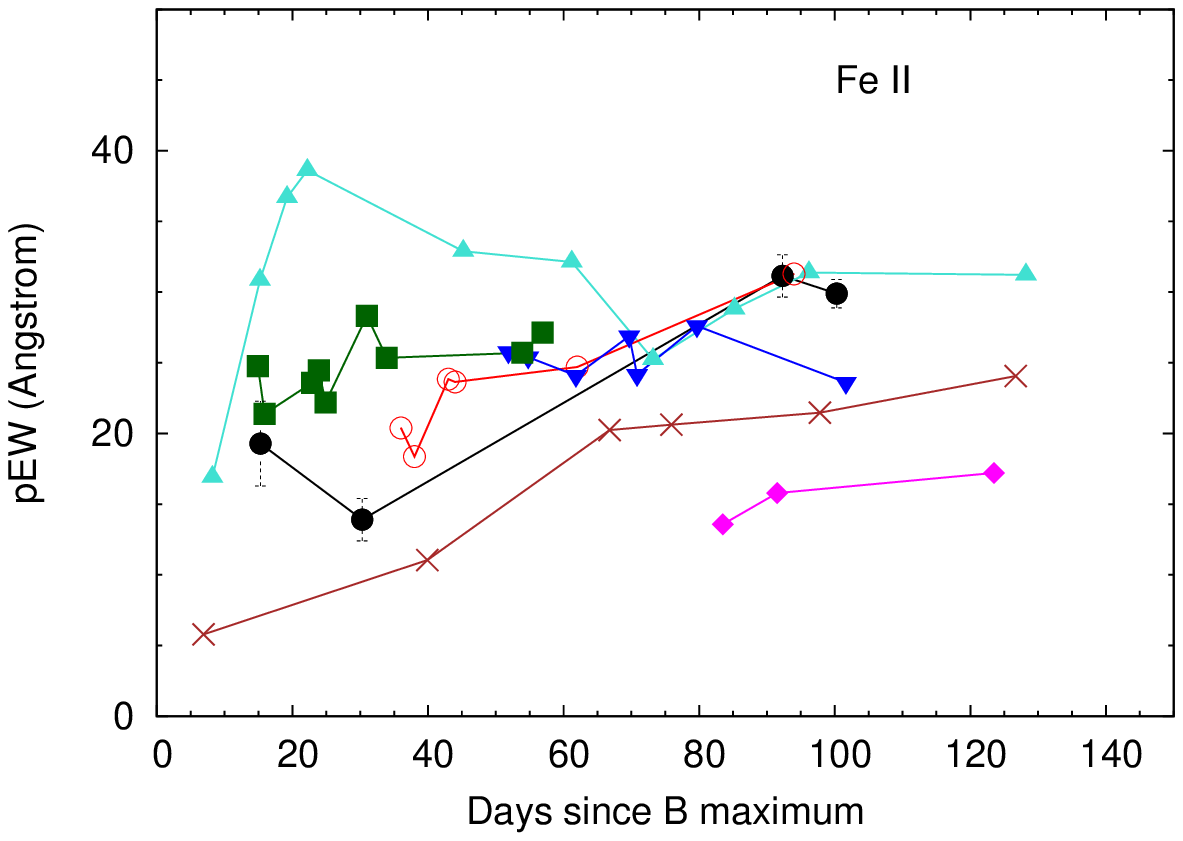} }
\subfloat{
        \label{}
        \includegraphics[width=0.45\textwidth]{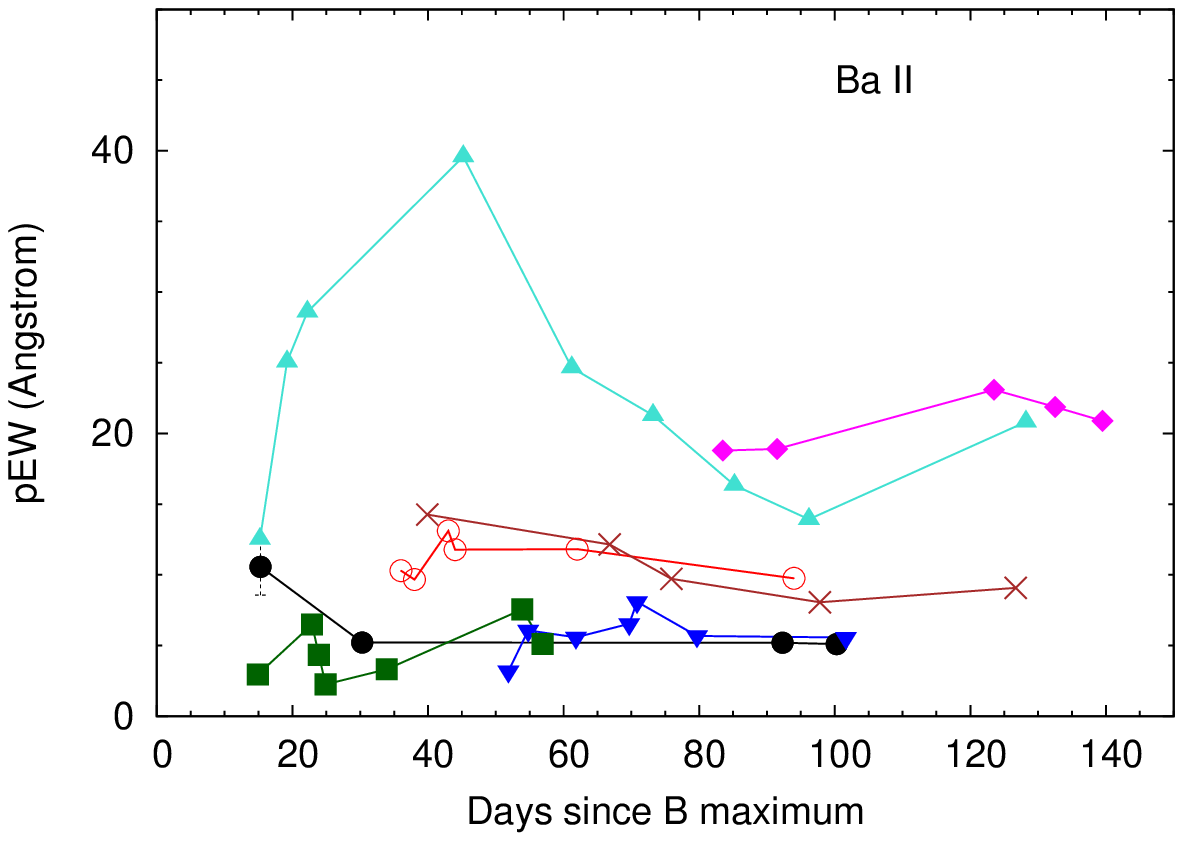} }
\caption{Pseudo equivalent widths of the absorption components of H$\alpha$ (top left panel), H$\beta$ (top right panel), Fe\,{\sc ii} $\lambda 5169$ (bottom left panel) and Ba\,{\sc ii} $\lambda 6142$ features of several 1987A-like SNe, compared to those of SN~2009mw.}
\label{pEWs}
\end{figure*}

In order to quantitatively compare the strength of some of the features in different SNe, we measured the pseudo equivalent widths (pEWs) of the absorption components of the H$\alpha$, H$\beta$, Fe\,{\sc ii}~$\lambda 5169$ and Ba\,{\sc ii} $\lambda 6142$ features. In Figure~\ref{pEWs} we compare these values. The pEWs of H$\alpha$ show large scatter among the different supernovae, while the pEWs of H$\beta$ seem to split into two groups at phase 40 days after explosion, SNe 2009mw, 1998A, 2000cb and 2006au having pEWs greater than 30~\AA, while the pEW values of SNe 1987A, 2006V and 2009E  are significantly smaller, less than 10~\AA~ at this epoch. We can detect similar duality in the pEWs of Ba\,{\sc ii}, in this case SNe 1987A and 2009E have greater values than the rest. The pEWs of the H\,{\sc i} lines of SN~1987A show an interesting behavior, the pEWs of both H$\alpha$ and H$\beta$ increase rapidly after explosion, reaching the peak at $19$ d and $8$ d after explosion, respectively, then decrease rapidly. \citet{87A_Balmerlines} have analysed this behavior and concluded that it is due to the temperature evolution  and the difference in density between the layers that produce H$\alpha$ and H$\beta$.
In our sample SNe 2000cb and 2006au seem to show similar behavior, though they're less well-sampled.

\begin{figure}
\includegraphics[width=84mm]{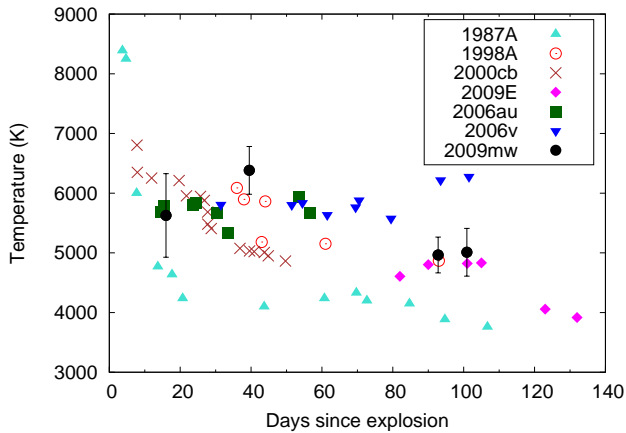}
\caption{The evolution of the temperature of SN~2009mw in comparison with other 1987A-like SNe. The temperatures were measured by fitting blackbody curves to the spectra, except for SN~2000cb the spectra of which are not flux calibrated, therefore the photometric SED was fitted with a blackbody curve.}
\label{temp}
\end{figure}

The first spectrum of SN~2009mw is early enough to indicate that it does not behave the same way, or, possibly, the maximum of the pEW of the Balmer lines are earlier than in the case of SN~1987A. We measured the temperatures of the SNe in our sample by fitting blackbody curves to their spectra (Fig.~\ref{temp}). The temperature of SN~1987A shows rapid decrease until about $20$~d after explosion. The early time temperatures of SN~2000cb are lower and the decrease is less steep, while in the case of the rest of the SNe we do not have sufficient observations to compare their behavior at this phase.
After $20$~d after explosion, all the SNe in out sample have significantly higher temperatures than SN~1987A, which is also reflected by the colour curves (Fig.~\ref{colour}).

The pEWs of the Fe\,{\sc ii} lines show some scatter at early phases but are quite similar for all the SNe except 2009E at around phase $70$ d after explosion. \citet{anderson_sneiimetallicity} found a correlation between the pEW of the Fe\,{\sc ii} features and the oxygen abundance of the host environment of SNe~II. Assuming, that this correlation exists for 1987A-like SNe, similar pEWs of Fe\,{\sc ii} lines would indicate that the objects in our sample emerged from similar environments (see also Sect.\ref{sec_host}). Figure~\ref{pEWs} also indicates, that SN~2009E is quite peculiar since the pEWs of almost all of its lines are low, except that of the Ba\,{\sc ii} feature, which is significantly higher than most of the other SNe.

\begin{figure}
\includegraphics[width=84mm]{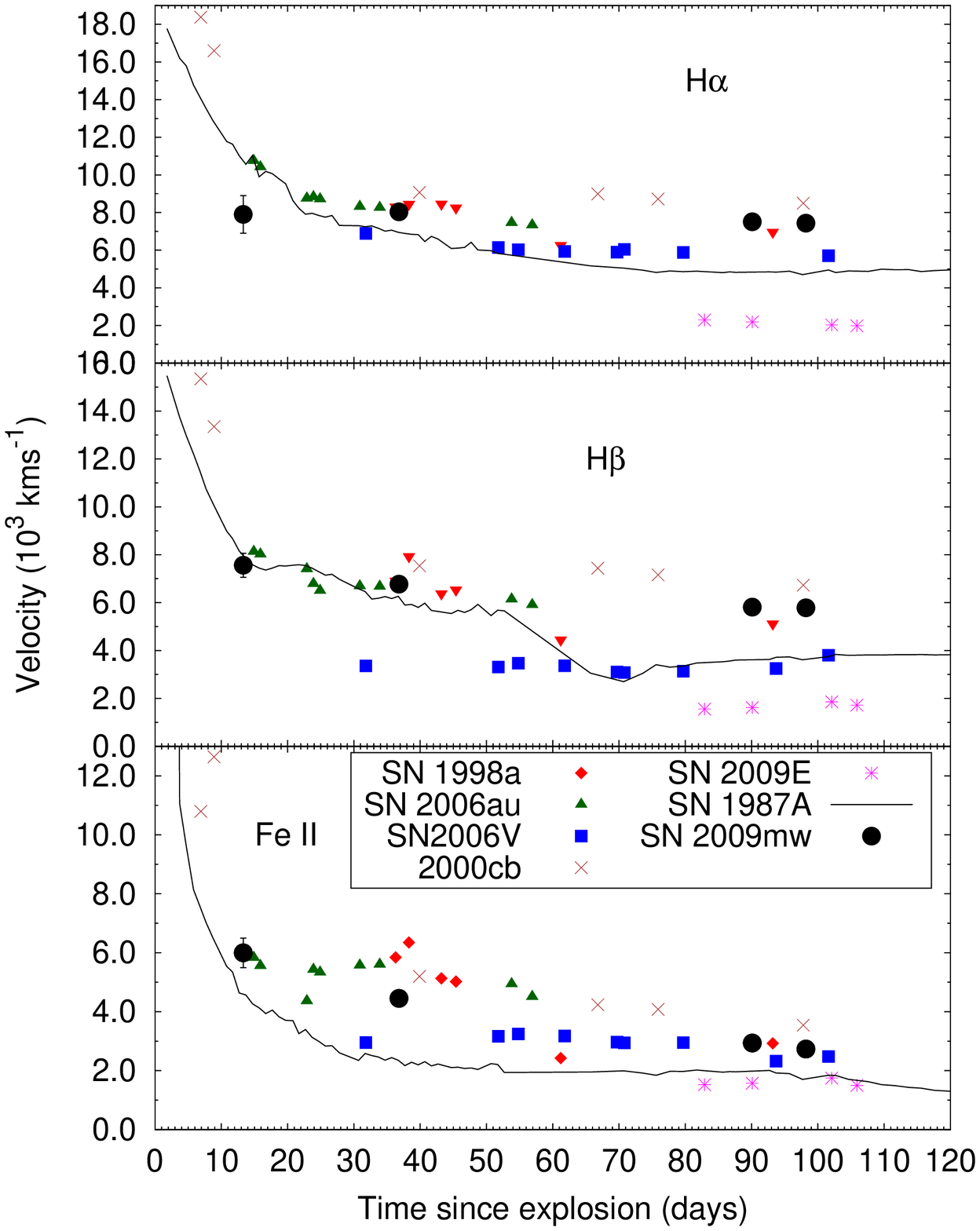}
\caption{The expansion velocities of SN~2009mw (black circles) in comparison with those of other 1987A-like SNe, measured from the Doppler-shift of the absorption minima of the H\,$\alpha$ (top), H\,{$\beta$} (middle) and Fe\,{\sc ii}~5169 (bottom) features.}
\label{vel_comp}
\end{figure}

We measured the Doppler-velocity of several lines in the spectra of SN 2009mw as well as in those of the other 1987A-like SNe. Fig.~\ref{vel_comp} compares the velocities measured from H$\alpha$, H$\beta$ and Fe\,{\sc ii} $\lambda$5169. These velocities show little differences, though during the first 10 days of evolution we only have data of SNe 1987A and 2000cb, and this is the period when the velocity changes more rapidly. At later phases the velocity curves are quite flat. SN 2009E had somewhat lower velocities than the others in this sample, while SN 2009mw had values similar to those of SNe 1998A and 2006au.

\begin{table}
 \centering
 \begin{minipage}{0.5\textwidth}
  \caption{Summary of the spectroscopic observations.}
  \label{logofsp}
  \begin{tabular}{@{}ccccc@{}}
  \hline
  \hline
Date & MJD & Phase$^a$ & Instrument set-up   \\
     & (days) & (days) &  \\
  \hline
25/12/2009 & 55190.3 & $-$69 & Gemini-S$+$GMOS $+$ R150  \\
17/01/2010 & 55213.8 & $-$45 & Gemini-S$+$GMOS $+$ R400  \\
12/03/2010 & 55267.1 & $+$8 & SOAR$+$GOODMAN $+$ 300 \\
20/03/2010 & 55275.2 & $+$16 & du Pont$+$WFCCD $+$ blue grism  \\
\hline
\end{tabular}
 \begin{tablenotes}
       \item[a]{$^a$ relative to the epoch of the $B$ maximum, i.e. $MJD=55258.8$}
     \end{tablenotes}
\end{minipage}
\end{table}

%##################### physical parameters
\section{Physical parameters}\label{sec_physicalparameters}

Comparing the observed parameters of a SN to those derived from semi-analytical and hydrodynamical modeling is a frequently used technique to estimate the physical properties of the progenitor and the explosion. Here we use a one dimensional Lagrangian LTE radiation hydrodynamic code \citep{bersten_model} for this purpose. We adopted double polytropic models as initial condition for our simulations. These parametric models preserve the typical density structure of blue supergiant (BSG) stars, which are commonly assumed to be the progenitors of 1987A-like SNe. While pre-SN models calculated from stellar evolution have a more clear physical grounding, especially if we are to connect the pre-SN mass with the mass of the star on the main sequence, we do not have access to a set of such models for BSGs. We note that in double polytropic models the pre-SN mass and radius are assumed to be independent parameters to be determined based on the comparison with the observations.
The explosion itself was simulated by injecting a certain amount of energy near the center of the progenitor object. We simultanously model the velocity curve, using the velocities determined from the Doppler-shift of the Fe\,{\sc ii} $\lambda$5169 line, and the bolometric luminosity curve due to the degeneracy between the progenitor mass and expansion velocities.

\begin{figure*}
\centering
\subfloat{
        \label{}
        \includegraphics[width=0.45\textwidth]{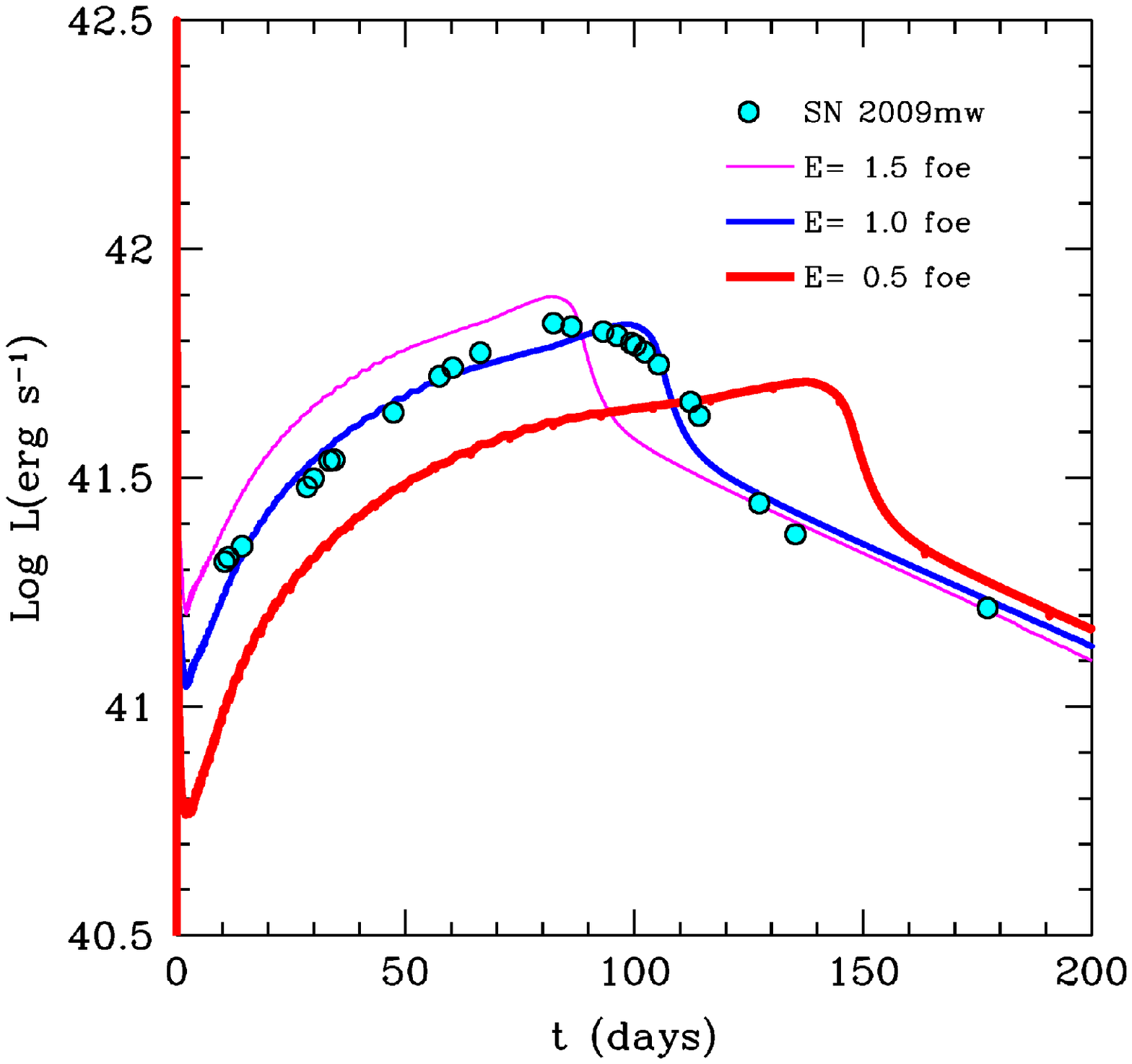} }
\subfloat{
        \label{}
        \includegraphics[width=0.45\textwidth]{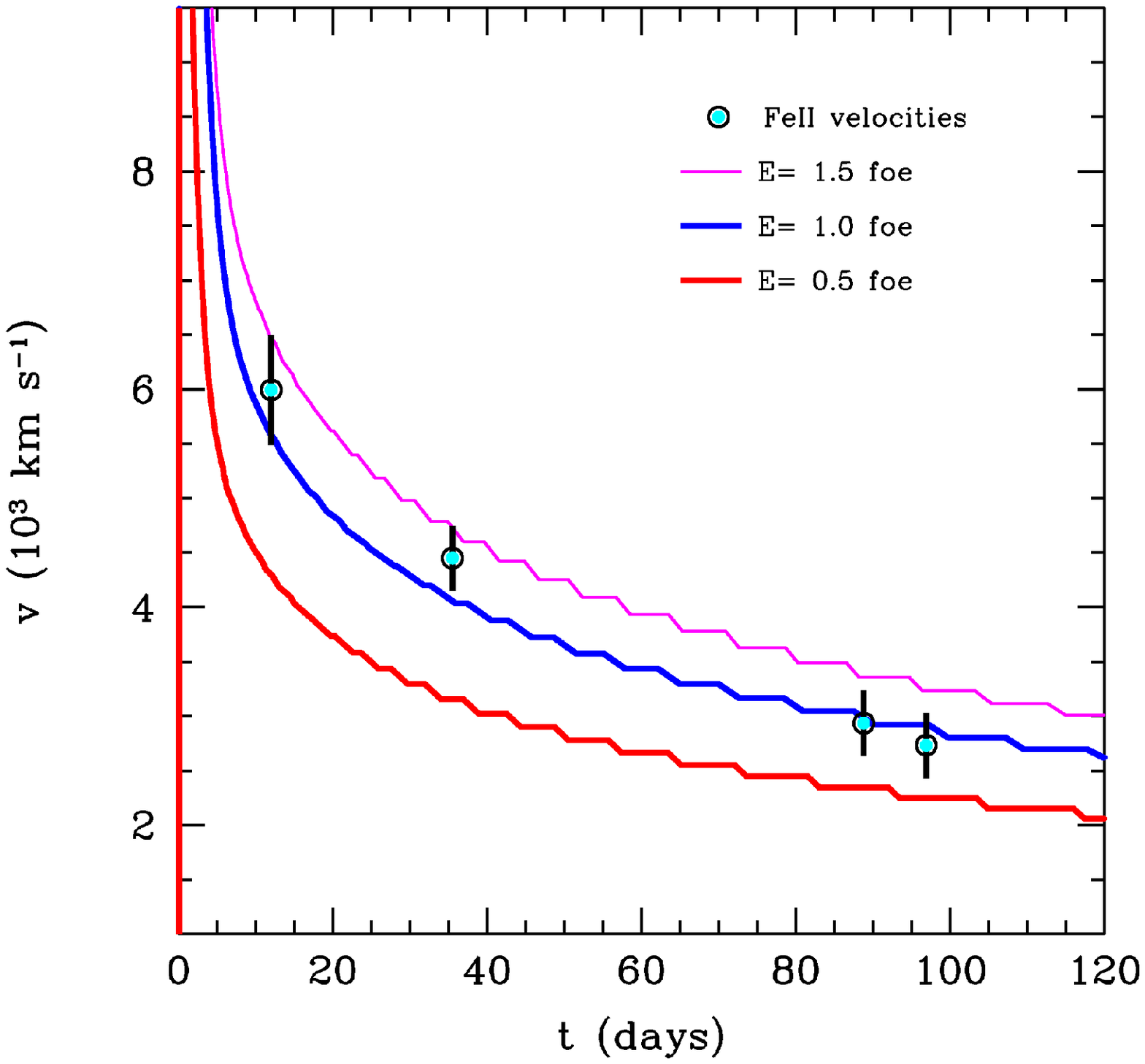} }\\
\subfloat{
        \label{}
        \includegraphics[width=0.45\textwidth]{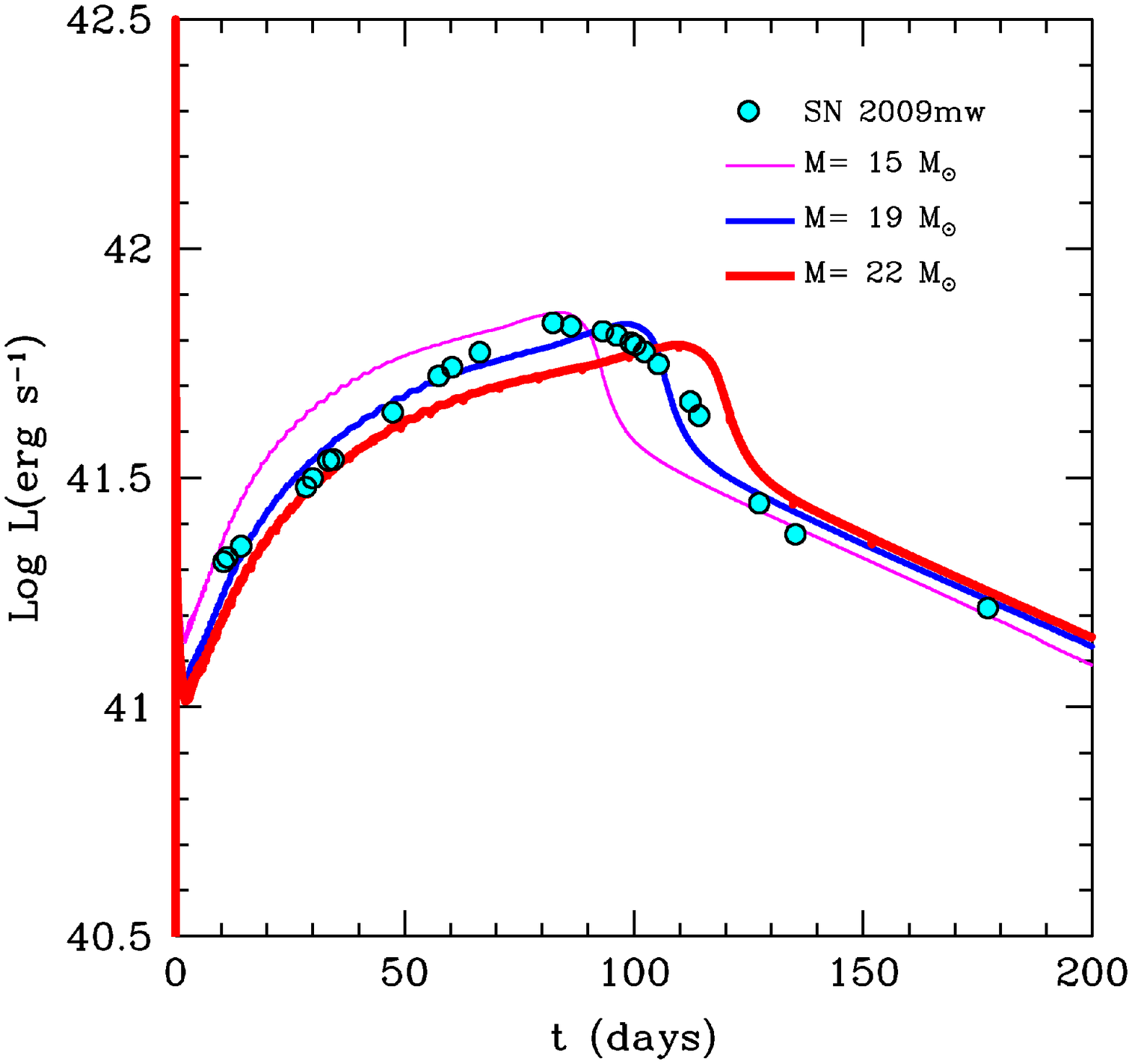} }
\subfloat{
        \label{}
        \includegraphics[width=0.45\textwidth]{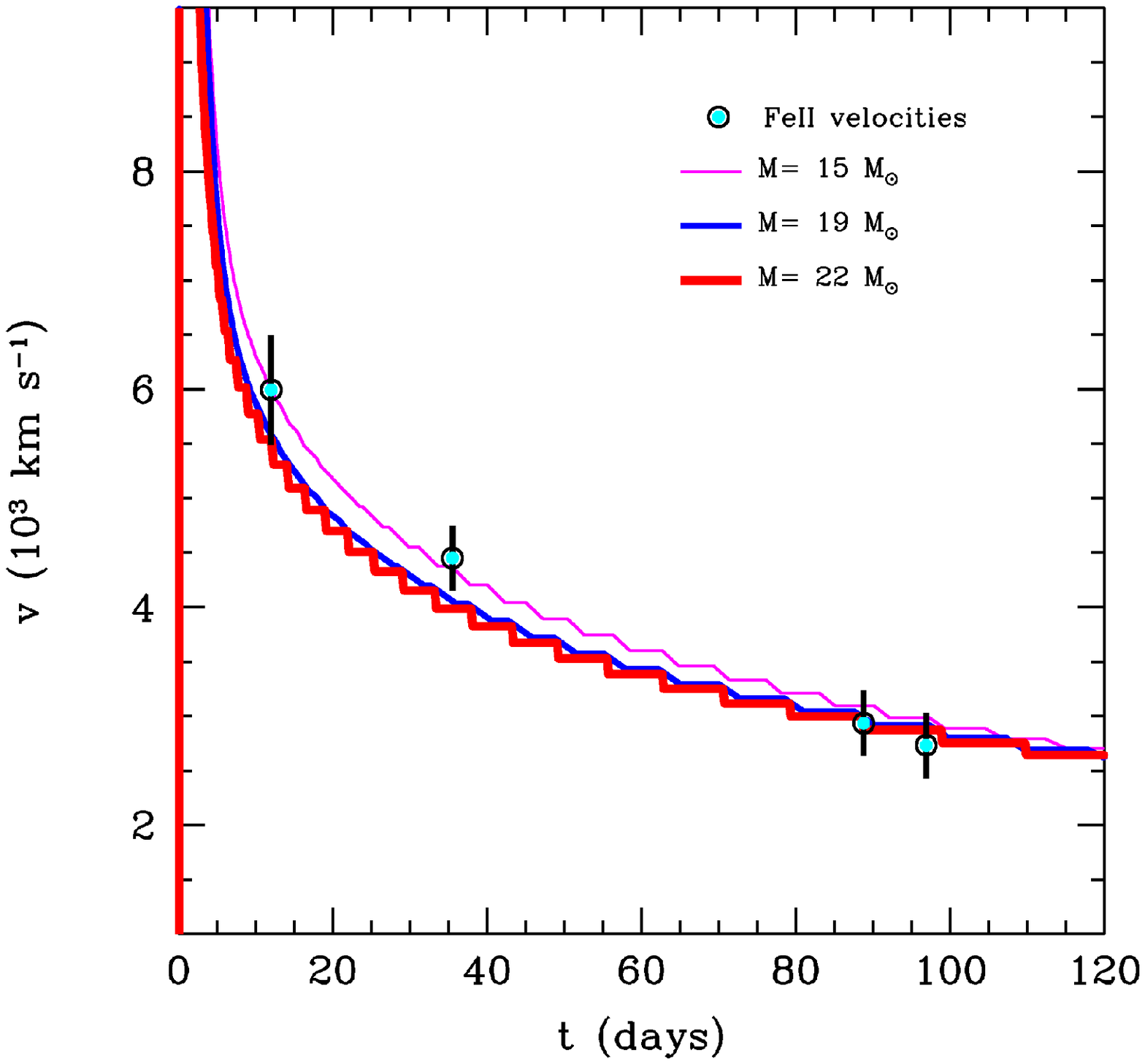} }
\caption{The bolometric luminosity (left panels) and the photospheric velocity (right panels) of SN~2009mw (points) compared to hydrodynamical models with different parameters. The blue line shown in each panel corresponds to the best-fitting model, {\sc M19R30E1Ni062}. The upper figures demonstrate the change of the model curves if we keep all parameters the same except the explosion energy, while the lower panels show the models with different ejecta mass.}
\label{lc_model}
\end{figure*}

Since we do not have UV or NIR data to directly calculate the bolometric luminosity, we followed the example of \citet{pastorello_09E}, by assuming that the bolometric correction is identical for SNe~1987A and 2009mw. We calculated $L_{BVRI,87A}$ using the same method as for SN 2009mw (Sec.\ref{sec_dist_bol}), while the bolometric luminosity curve was adopted from the literature \citep{catchpole_87A_I, catchpole_87A_II}. The determined bolometric correction for SN~1987A allowed us to calculate the $L_{bol,09mw}$ curve, which is shown in Figure~\ref{lc_model}.

 In the same figure we also present the best-fitting model, which corresponds to a progenitor radius of $30~R_{\odot}$, an explosion energy of $1$~foe, a $^{56}$Ni mass of $0.062~M_{\odot}$, and a pre-SN mass of $19\,{\rm M_{\odot}}$, assuming that a compact remnant of $1.5\,{\rm M_{\odot}}$ was formed during the explosion (model {\sc M19R30E1Ni062}). In order to find this model, several calculations with different masses, radii, energies, and radioactive material have been performed. Fortunately, each parameter affects the resulted light curve in a particular way. Even if there is degeneracy between the explosion energy and the pre-SN mass, this can be removed by considering the expansion velocities. This is demonstrated in Figure~\ref{lc_model}, where the sensitivity of the model to the explosion energy and the pre-SN mass (upper and lower panels, respectively) is shown. In this figure models with different explosion energies ($E= 0.5, 1,\,{\rm and}\,1.5$~foe)
 and masses ($M=15, 19,\,{\rm and}\,22\,{\rm M_{\odot}}$) are presented. We kept the other parameters identical to those of model  {\sc M19R30E1Ni062}. As shown, changing the pre-SN mass has only a small effect on the resulting velocity, it is the explosion energy that mainly regulates the photospheric velocity evolution. After setting the energy from this comparison, we varied the rest of the parameters until we found the best model that describes the observed evolution. We also found that changing the progenitor radius mainly affects the early phase of the light curve, while the $^{56}$Ni mass determines the tail luminosity.

The parameters of some of the 1987A-like SNe have been determined with similar techniques, Table~\ref{physpar} summarizes these results. It shows that the progenitors of these objects are all compact stars, with a progenitor radius less than $100\,{\rm R_{\odot}}$. The ejecta mass is close to $20\,{\rm M_{\odot}}$ in all cases (though one of the models for SN~1987A found smaller values). The variation of the explosion energy is the range of a few foe, and the nickel mass changes in the range between  $0.04-0.13\,{\rm M_{\odot}}$.

\begin{table}
 \centering
 \begin{minipage}{0.5\textwidth}
  \caption{Comparison of the physical parameters of the progenitor and explosion of 1987A-like SNe. The table is ordered by the brightness of the SNe, the faintest one being on the top.}
  \label{physpar}
  \begin{tabular}{@{}lccccc@{}}
  \hline
  \hline
SN & $E$ & $M_{ejecta}$ & $R$ & $^{56}Ni$ &  Ref.\\
& (foe) & (M$_{\odot}$) & (R$_{\odot}$) & (M$_{\odot}$) &\\
\hline
2009E  & 0.6 & 19.0 & 100 & 0.040 & 1 \\
2009mw & 1.0 & 17.5 & 30 & 0.062 &  this paper \\
1987A & 1.1 & 11.8 & 33 & 0.078 & 2 \\
1987A & 1.6 & 18.0 & 72 & 0.075 &  3 \\
2000cb & 4.4 & 22.3 & 35 & 0.083 & 4 \\
1998A & 5.6 & 22.0 & $\leq 86$ & 0.11 &  3 \\
2006au & 3.2 & 19.3 & 90 & $\leq 0.073$ & 2\\
2006V & 2.4 & 17.0 & 75 & 0.127 & 2 \\
\hline
\end{tabular}
 \begin{tablenotes}
      \item[a]{References: (1) \citet{pastorello_09E}, (2) \citet{taddia_87alike}, (3) \citet{pastorello_98a}, (4) \citet{utrobin_2000cb}.}
      \item[b]{Note, that \citet{pastorello_98a} used the semianalytic code of \citet{zampieri2003} to obtain the parameters of SNe 1987A and 1998A, \citet{pastorello_09E}  applied the radiation-hydrodynamics code of \citet{pumo2010} and \citet{pumo2011}, \citet{taddia_87alike} used the semi-analytic model of \citet{imshennik_1992}, while \citet{utrobin_2000cb} used their own hydrodynamical model}.
     \end{tablenotes}
\end{minipage}
\end{table}

%#######################3 host
\section{host environment}\label{sec_host}

\citet{taddia_87Alike_host} studied the environments of 1987A-like SNe, and found that these objects are located either on the outer parts of bright, metal-rich galaxies or in faint, metal-poor hosts, and that on average these SNe emerge from populations with slightly lower metallicity than that of the hosts of normal SNe II-P.

\begin{figure}
\includegraphics[width=84mm]{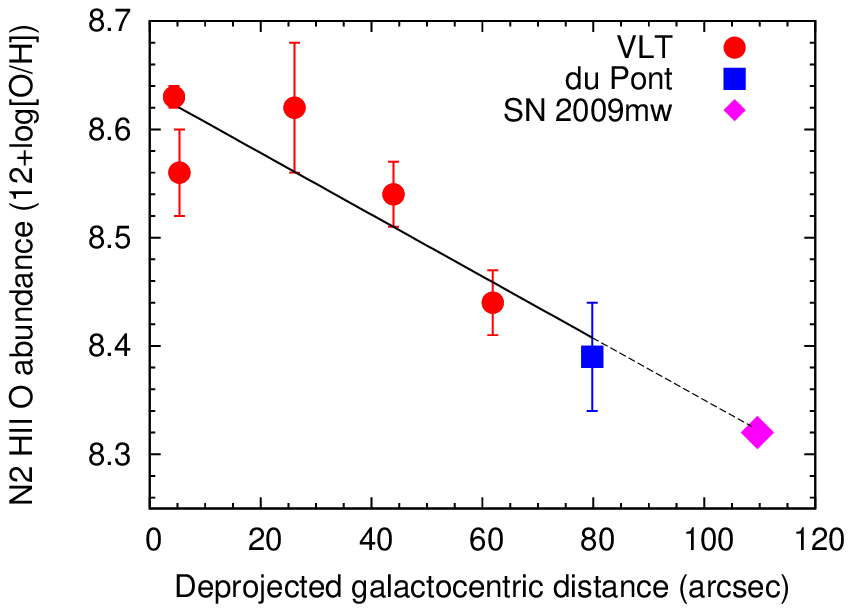}
\caption{Using multiple measuremets at different deprojected distances, we estimated the oxigen abundance at the position of SN~2009mw via extrapolation. We used a long slit spectrum taken by Anderson et al. (in prep.) with VLT to extract spectra of several H\,{\sc ii} region (red circles), as well as our spectrum of SN~2009mw taken with the du Pont telescope to take measurement close to the site of the SN (blue square). The extrapolation yielded $N2=8.32 \pm 0.08$~dex, a subsolar value for the position of SN~2009mw (magenta diamond).}
\label{Z_grad}
\end{figure}

The host galaxy of SN~2009mw, ESO 499-G005 is a type SAB(s)c galaxy, with the absolute brightness of $M_B=-19.7$~mag. The SN is located in the outskirts of the galaxy (see Fig.~\ref{picture}), the ratio between its deprojected position and $r_{25}$ is $d_{\rm SN}/r_{25}=1.4$. We were able to extract a spectrum of an H\,{\sc ii} region close to the position of the SN from the spectrum taken with the du Pont telescope (Sec.~\ref{sec_spectroscopy}). Using the relations of \citet{marino_metallicity} we determined the oxigen abundance ($12+\log(O/H)$) as $N2=8.39 \pm 0.05$~dex (note that the systematic error of the diagnostic is $0.16$~dex).
 As part of a project to determine the metallicity of parent populations of SNe, there were VLT$+$FORS2 spectrum taken of this galaxy, aiming for another SN it hosted, SN~2008H (Anderson et al., in prep.). SN 2008H was located in the center of the galaxy, that is the position where the slit was placed.  We were able to extract spectra at several positions along the slit and used the relations of \citet{marino_metallicity} to determine the oxygen abundance. The derived metallicity is solar-like close to the center of the galaxy and decreases along the deprojected distance from it. None of our measurements extend as far as the deprojected distance of SN~2009mw, but by extrapolating the gradient we can estimate the oxygen abundance at the distance of SN 2009mw as $N2=8.32 \pm 0.08$~dex, a somewhat sub-solar value (see Fig.~\ref{Z_grad}).

%##########################3 summary
\section{Discussion}\label{sec_discussion}

In this paper we present optical photometry and spectroscopy of SN~2009mw. This SN belongs to the small subgroup of 1987A-like SNe, which are Type II SNe characterized by a light curve which keeps rising for about 3 months after explosion. The current sample of 1987A-like SNe is very small, which originates from the fact that they are intrinsically rare. To date, data of less than 18 objects has been published, and only a few of them have multiband photometric and spectroscopic data available.

We obtained photometric follow-up of SN~2009mw using $BVRI$ and $g'r'i'z'$ bands. Comparing the shape of the light curves to those of SNe~1987A, 1998A, 2006V, 2006au and 2009E we found that they are remarkably similar, while SN~2000cb has a light curve that is quite different, the slow brightening and late maximum in $B$ band is missing. The colour of SN~2009mw is bluer than that of SNe~1987A and 2000cb, and slightly redder than that of SNe~2006V and 2006au, more alike to that of SNe~2009E and 1998A. The absolute brightness of SN~2009mw is close to that of SNe~1987A and 2009E and fainter than those of SNe 1998A, 2000cb, 2006V and 2006au.

The spectra of 1987A-like SNe are similar to those of SNe II-P. The spectra of the objects in our sample evolve similarly, showing variations mostly in the strengths of the Ba\,{\sc ii} and H\,$\beta$ lines. In this aspect, the spectra of SN~2009mw are more akin to those of SNe~1998A and 2006au.

We estimated the pysical properties of the progenitor and the explosion via hydrodynamical modelling, yielding an explosion energy of $1$~foe, a pre-SN mass of $19\,{\rm M_{\odot}}$, a progenitor radius of $30\,{\rm R_{\odot}}$, and a $^{56}$Ni mass of $0.062\,{\rm M_{\odot}}$. These values are similar to those obtained for SN~1987A, indicating that the progenitor of SN~2009mw was a blue supergiant star, as was the case for SN~1987A. The physical parameters of several 1987A-like SNe have been estimated via modeling, all suggesting compact progenitors.

We examined the metallicity of the host environment of SN~2009mw, and found a somewhat sub-solar value for the oxigen abundance, aligned with the findings of \citet{taddia_87Alike_host}, namely that 1987A-like SNe emerge from environments with slighty lower metallicty than those of SNe II-P.

Our comparison of SN~2009mw to other 1987A-like SNe shows, that while some of their parameters show variations, others such as their spectra, progenitors and environments are quite similar. We need a greater sample of these objects in order to understand that, for example, why the light curves of SNe 2000cb and 2004ek stand out so much, while their progenitors are similar to the rest. 1987A-like SNe seem to emerge from BSG stars located in environments with slightly sub-solar metallicity, though the processes that produce and explode these stars are still debated.

%#########################3 acknowledgement
\section*{Acknowledgement}

We thank N. Morrell and the {\em Carnegie Supernova Project} (CSP) for the spectrum of SN~2009mw taken with the du Pont telescope. We thank the referee for the useful comments that helped to improved this manuscript.

KT was supported by CONICYT through the FONDECYT grant 3150473.
KT, GP and MH received support from the Ministry of Economy, Development, and Tourism’s Millennium Science Initiative through grant IC12009, awarded to the Millennium Institute of Astrophysics, MAS. MS gratefully acknowledges support from the  Danish Agency for Science and Technology and Innovation realized through a Sapere Aude Level 2 grant.

Based on observations obtained at the Gemini Observatory under the program ID GS-2009B-Q-9. The Gemini Observatory is operated by the Association of Universities for Research in Astronomy, Inc., under a cooperative agreement with the NSF on behalf of the Gemini partnership: the National Science Foundation (United States), the National Research Council (Canada), CONICYT (Chile), the Australian Research Council (Australia), Minist\'{e}rio da Ci\^{e}ncia, Tecnologia e Inova\c{c}\~{a}o (Brazil) and Ministerio de Ciencia, Tecnolog\'{i}a e Innovaci\'{o}n Productiva (Argentina).

Spectra of SNe~1987A, 1998A, and 2009E were downloaded from WISeREP \citep[Weizmann Interactive Supernova data REPository, \url{http://wiserep.weizmann.ac.il/}][]{wiserep}. This research has made use of the NASA/IPAC Extragalactic Database (NED) which is operated by the Jet Propulsion Laboratory, California Institute of Technology, under contract with the National Aeronautics and Space Administration.
%########################## biblio
\bibliographystyle{mn2e}
\bibliography{SN2009mw_takats.bib}

\end{document}